\documentstyle[aps,floats,graphicx]{revtex}

\begin{document}
\draft
\renewcommand{\textfraction}{0.0}

\wideabs {
\title {
   The analysis of single crystal diffuse scattering using the Reverse Monte 
   Carlo method: Advantages and problems }

\author {Th. Proffen and T. R. Welberry}

\address{Research School of Chemistry, 
	 The Australian National University, \\
         Canberra, ACT 0200, Australia }
\date{\today}
\maketitle


\begin{abstract}
The scattering from crystals can be divided into two parts: Bragg 
scattering and diffuse scattering.  The analysis of Bragg diffraction data 
gives only information about the {\it average} structure of the crystal.  
The interpretation of diffuse scattering is in general a more difficult 
task.  A recent approach of analysing diffuse scattering is based on the 
Reverse Monte Carlo (RMC) technique.  This method minimises the difference 
between observed and calculated diffuse scattering and leads to {\it one} 
real space structure consistent with the observed diffuse scattering.  \par

The first example given in this paper demonstrates the viability of the RMC 
methods by refining diffuse scattering data from simulated structures 
showing known occupational and displacement disorder.  As a second example, 
results of RMC refinements of the diffuse neutron- and X-ray scattering of 
stabilised zirconia (CSZ) are presented.  Finally a discussion of the RMC 
method and an outlook on further developments of this method is given.
\end{abstract}

}


\section{Introduction}

Crystal structure analysis based on Bragg diffraction data reveals only 
information about the {\it average} crystal structure, such as atomic 
positions, thermal ellipsoids and site occupancies.  Any departure from the 
strictly long-range-ordered {\it average} structure of the crystal gives 
rise to diffuse scattering containing information about static or thermal 
disorder within the studied material.  Since many properties like optical 
properties, hardness, ionic conductivity are governed by structural 
disorder, the determination of the defect structure based on diffuse 
scattering data can reveal important information about the studied 
material.  The development of area detectors (proportional counters, image 
plates, CCD) in recent years has enormously increased the capability for 
measuring generally weak diffuse neutron and X-ray scattering data.  
However, the interpretation and analysis of diffuse scattering remains a 
generally difficult task.  The availability of modern (super)computers 
opens a wide area of computer simulation techniques to aid the analysis of 
diffuse scattering.  An overview of traditional approaches and the analysis 
of diffuse scattering via computer simulations can be found in 
\cite{webu94}, further general information about disorder diffuse 
scattering can be found in numerous review articles 
\cite{we85,ja87,jafr93,fr95,fr97,webu95}.  We want to focus in this paper 
on the analysis of diffuse single crystal scattering using the RMC method 
developed by McGreevy and Pusztai \cite{mcpu88}.  \par

The RMC refinement technique minimises the difference between observed and 
calculated diffuse scattering intensities as a function of the positions 
and occupancies of the atomic sites in the model crystal.  Although the RMC 
method is known for about 10 years, the application of RMC to diffuse {\it 
single crystal} scattering data was reported just recently in a neutron 
diffraction study of the diffuse scattering of ice {\it Ih} 
\cite{nikemc95}.  The advantage of the RMC method is the fact, that it is a 
{\it model free} method to analyse diffuse scattering, i.e.  no assumptions 
about the particular disordered structure under investigation have to be 
made.  In general a RMC simulation gives {\it one} real-space structure 
consistent with the experimental data.  The remaining difficulty is the 
interpretation of the resulting structure.  \par

The aim of this paper is to give an introduction into the RMC simulation 
technique for the analysis of diffuse {\it single crystal} scattering and 
to discuss the advantages and difficulties of the RMC method.


\section{The RMC method}

In general Monte Carlo methods can be described as statistical simulation 
methods involving sequences of random numbers to perform the simulation.  
In the past several decades this simulation technique based on the 
algorithm developed by Metropolis \cite{merorotete53} has been used to 
solve complex problems in nuclear physics, quantum physics, chemistry as 
well as for simulations of e.g.  traffic flow or econometrics.  The name 
{\it Monte Carlo} was coined during the Manhattan Project in World War II, 
because of the similarity of statistical simulation to games of chance, and 
because the capital of Monaco was a center of gambling.  In this analogy 
the 'game' is a physical system and the scientist might 'win' a solution 
for his particular problem.  An excellent application for this kind of 
statistical method is the study of diffuse scattering and subsequently the 
solution of the underlying defect structure.  One possible approach is the 
(direct) Monte Carlo (MC) modeling of a defect structure from a given set 
of near-neighbour interaction energies \cite{webu94}.  The same basic 
algorithm is used for RMC simulations to minimise the difference between 
calculated and measured diffuse scattering as described in the following 
section.

\subsection{How does it work ?}

As described in the introduction, the aim of the RMC simulation process is 
to minimise the difference between observed and calculated diffraction 
pattern.  As a first step, the scattered intensity is calculated from the 
chosen crystal starting configuration and a goodness-of-fit parameter 
$\chi^{2}$ is computed.

\begin{equation}
	\chi^{2} = \sum_{i=1}^{N} 
	           \frac{ ( I_{e}({\bf h}_{i}) - I_{c}({\bf h}_{i})) ^{2}}
	                {\sigma^{2}}
	\label{eq1}
\end{equation}

The sum is over all measured data points ${\bf h}_{i}$, $I_{e}$ stands for 
the experimental and $I_{c}$ for the calculated intensity.  The RMC 
simulation proceeds with the selection of a random site within the crystal.  
The system variables associated with this site, such as occupancy or 
displacement, are changed by a random amount, and then the scattered 
intensity and the goodness-of-fit parameter $\chi^{2}$ are recalculated.  
The change $\Delta\chi^{2}$ of the goodness-of-fit $\chi^{2}$ before and 
after the generated move is computed.  Every move which improves the fit 
($\Delta\chi^{2} < 0$) is accepted.  'Bad' moves worsening the agreement 
between observed and calculated intensity are accepted with a probability 
of $P=\exp(-\Delta\chi^{2}/2)$.  As the value of $\Delta \chi^{2}$ is 
proportional to $1 / \sigma^{2}$, the value of $\sigma$ has an influence on 
the amount of 'bad' moves which will be accepted.  Obviously there are two 
extremes: For very large values of $\sigma$, the experimental data are 
ignored ($\chi^{2} \approx 0$) and with very small values of $\sigma$ the 
fit ends up in the local minimum closest to the starting point, because 
there is a negligible probability for 'bad' moves.  The parameter $\sigma$ 
acts like the temperature $T$ in 'normal' MC simulations.  The RMC process 
is repeated until $\chi^{2}$ converges to its minimum.  \par

The result of a successful RMC refinement is {\it one} real space structure 
which is consistent with the observed diffuse scattering data.  In order to 
exclude chemically implausible resulting structures additional constrains, 
e.g.  minimal allowed distances between atoms, may be introduced.

\subsection{RMC software}

\begin{figure}[!htbp]
\centering
\includegraphics[width=2.8in]{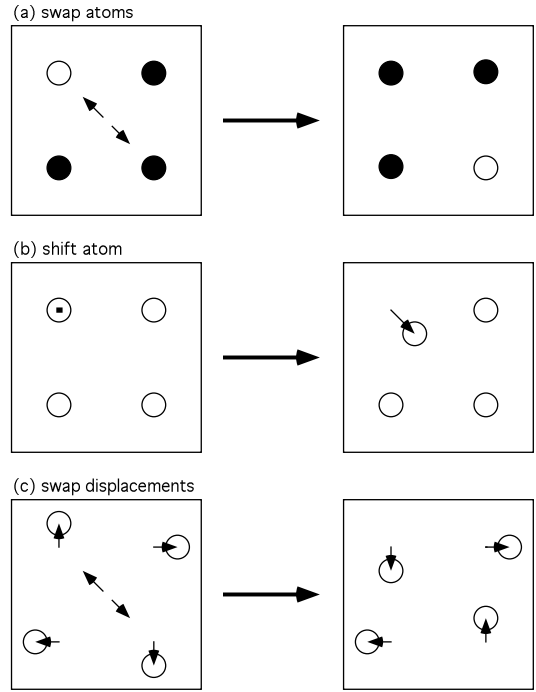}
\caption{ Illustration of the different RMC modes: (a) swap atoms, (b) shift 
	  atom and (c) swap displacements from average position.}
\label{fig1}
\end{figure}

The program used for the RMC refinements presented in this paper is {\it 
DISCUS} \cite{prne97}.  For practical use it is necessary to include a 
scaling factor $f$ and a background parameter $b$ in the previous 
definition of the goodness-of-fit $\chi^{2}$.  A weight $w({\bf h})$ is 
included as well.  {\it DISCUS} allows the user to choose a particular 
weighting scheme or read weights from a separate input file.  The 
definition of $\chi^{2}$ becomes:

\begin{equation}
	\chi^{2} = \sum_{i=1}^{N} \frac { w({\bf h}_{i}) [ I_{e}({\bf h}_{i}) - 
	           (f \cdot I_{c}({\bf h}_{i}) + b) ] ^{2}} { \sigma^{2} }
	\label{eq2}
\end{equation}

As before $I_{e}({\bf h}_{i})$ stands for the measured intensity at the 
reciprocal point ${\bf h}_{i}$, and $I_{c}({\bf h}_{i})$ is the calculated 
intensity in that point.  The summation is over all N experimental data 
points.  Three different ways to calculate the scale $f$ and background $b$ 
are implemented.  First the user can define fixed values for both: $f = 
f_{0}, b = b_{0}$.  Secondly, the background can be set to a fixed value $b 
= b_{0}$ and the scaling factor $f$ is computed according to:

\begin{equation}
    f = \frac {\sum\limits_{i=1}^{N} w({\bf h}_{i}) I_{e}({\bf h}_{i}) 
                                             I_{c}({\bf h}_{i}) 
	   - b_{0} \sum\limits_{i=1}^{N} w({\bf h}_{i}) I_{c}({\bf h}_{i})}
	          {\sum\limits_{i=1}^{N} w({\bf h}_{i}) I_{c}^{2}({\bf h}_{i}) }
    \label{eq3}
\end{equation}
  
Alternatively both values $f$ and $b$ can be refined during the RMC 
refinement. Equation (\ref{eq4}) shows the corresponding definitions.

\begin{figure*}[!htbp]
\begin{eqnarray}
	f & = & \frac{\sum\limits_{i=1}^{N} w({\bf h}_{i})
	              \sum\limits_{i=1}^{N} w({\bf h}_{i}) 
	                                    I_{e}({\bf h}_{i})I_{c}({\bf h}_{i})    
	            - \sum\limits_{i=1}^{N} w({\bf h}_{i}) I_{e}({\bf h}_{i}) 
	              \sum\limits_{i=1}^{N} w({\bf h}_{i}) I_{c}({\bf h}_{i})}
	           {  \sum\limits_{i=1}^{N} w({\bf h}_{i})
	              \sum\limits_{i=1}^{N} w({\bf h}_{i}) I_{c}^{2}({\bf h}_{i})
	     - \left (\sum\limits_{i=1}^{N} w({\bf h}_{i}) 
	                                    I_{c}({\bf h}_{i}) \right ) ^{2} }
	\nonumber  \\
	b & = & \frac{\sum\limits_{i=1}^{N} w({\bf h}_{i}) I_{e}({\bf h}_{i}) 
	    - f \cdot \sum\limits_{i=1}^{N} w({\bf h}_{i}) I_{c}({\bf h}_{i})}
	             {\sum\limits_{i=1}^{N} w({\bf h}_{i})}
	\label{eq4}
\end{eqnarray}
\end{figure*}

The parameters $f$ and $b$ are computed in each RMC cycle and have usually 
large starting values as long as there are big differences between 
calculated and observed data.  After every RMC move the resulting 
scattering intensity and the $\chi^{2}$ value is calculated.  In order to 
save computing time only the contribution of the modified atoms to the 
scattering is calculated.  The difference $\Delta \chi^{2} = \chi_{old}^{2} 
- \chi_{new}^{2}$ is taken to decide if the move will be accepted or not as 
described in the previous section.  The program calculates separate scaling 
factors and background parameters for every used plane of experimental 
data.  \par

The program {\it DISCUS} is capable of modeling occupational as well as 
displacement disorder and so far we have called each crystal modification 
simply RMC move.  In practice we use a mode ('switch-atoms') of simulation 
in which occupational disorder is modeled by swapping two different 
randomly selected atoms (Fig. \ref{fig1}).  This procedure forces the relative 
abundances of the different atoms within the crystal to be constant.  It 
should be noted, that vacancies are treated as an additional atom type 
within the program {\it DISCUS}.  The introduction of displacement disorder 
is realised in two different ways.  In the first method a randomly selected 
atom is displaced by a random Gaussian distributed amount ('shift', Fig.  
\ref{fig1}b).  Alternatively the displacement variables associated with two 
different randomly selected atoms are interchanged ('switch-displacements', 
Fig. \ref{fig1}c).  The latter method has the advantage that the overall 
mean-square displacement averages for each atom site can be introduced into 
the starting model and these will remain constant throughout the 
simulation.  Additional information about the program {\it DISCUS} can be 
found on the World-Wide-Web \cite{discuswww}.


\section{Examples}
\subsection{RMC test refinements \label{test}}

In this first example, we test the viability of the RMC simulation
technique for systems showing occupational and displacement disorder
in combination. First a disordered structure with given correlation
parameters and displacements is created and the diffuse intensity
calculated. This intensity is used as input for the RMC refinement.
Finally the resulting structure is compared to the expected disordered
structure. Details about the complete series of test simulations
can be found in \cite{prwe97}.\par

\begin{table}[htbp]
\begin{tabular}{l r@{.}l r@{.}l r@{.}l r@{.}l}
  & 
  \multicolumn{2}{c}{{\it Input}} & 
  \multicolumn{2}{c}{{\it Run A}} & 
  \multicolumn{2}{c}{{\it Run B}} & 
  \multicolumn{2}{c}{{\it Run C}}\\
  \hline
  $c_{10}$      &  -0&203  &  0&188  & -0&076  &  -0&124 \\
  $c_{11}$      &   0&523  &  0&290  &  0&412  &   0&465 \\
  $d_{zr-zr}$ [\AA] &   5&05(12) & 5&02(9)  & 
                        5&02(8)  & 5&03(6)  \\
  $d_{zr-vac}$[\AA] &   4&89(9)  & 4&95(11) & 
                        4&97(10) & 4&94(10) \\
  $R$           &   \multicolumn{2}{c}{-} &  39&3 \% &  39&2 \% &  7&7 \% \\
\end{tabular}
\caption{\label{tab1}Input structure and results of RMC test 
          refinements}
\protect
\end{table}

The simulated structure used to calculate the 'experimental' data for the 
RMC refinements was a 2D square-symmetric crystal with a size of 50x50 unit 
cells, one atom (Zr) at (0,0,0) with an occupancy of 0.83 and a lattice 
constant of a=5\AA.  We chose the defect 'test' structure to consist of 
preferred vacancy pairs in $<$11$>$ direction and a subsequent relaxation 
of the surrounding atoms towards the vacancy.  The vacancy ordering can be 
described using the correlation coefficient $c_{ij}$ which is defined as:

\begin{equation}
	c_{ij} = \frac {P_{ij} - \theta^{2}} { \theta (1 - \theta)}
	\label{eq5}
\end{equation} 

$P_{ij}$ is the joint probability that both sites $i$ and $j$ are occupied 
by the same atom type and $\theta$ is its overall occupancy.  Negative 
values of $c_{ij}$ correspond to situations where the two sites $i$ and $j$ 
tend to be occupied by {\it different} atom types while positive values 
indicate that sites $i$ and $j$ tend to be occupied by the {\it same} atom 
type.  A correlation value of zero describes a random distribution.  The 
maximum negative value of $c_{ij}$ for a given concentration $\theta$ is 
$-\theta/(1-\theta)$ ($P_{ij}=0$), the maximum positive value is +1 
($P_{ij}=\theta$).  We will refer to the correlation coefficient of nearest 
neighbours in $<$10$>$ direction as $c_{10}$ and in $<$11$>$ direction as 
$c_{11}$.  The achieved vacancy concentration $\theta$ and correlation 
values $c_{10}$, $c_{11}$ are listed in Table \ref{tab1}.  The resulting 
structure is characterised by a positive correlation $c_{11}$ and a 
negative value for $c_{10}$ close to its maximum negative value of 
$-\theta/(1-\theta) = -0.205$.  In other words, the vacancy ordering is 
given by preferred $<$11$>$ vacancy pairs and avoided $<$10$>$ pairs. \par

\begin{figure*}[!htbp]
\centering
\includegraphics[angle=270,width=5.6in]{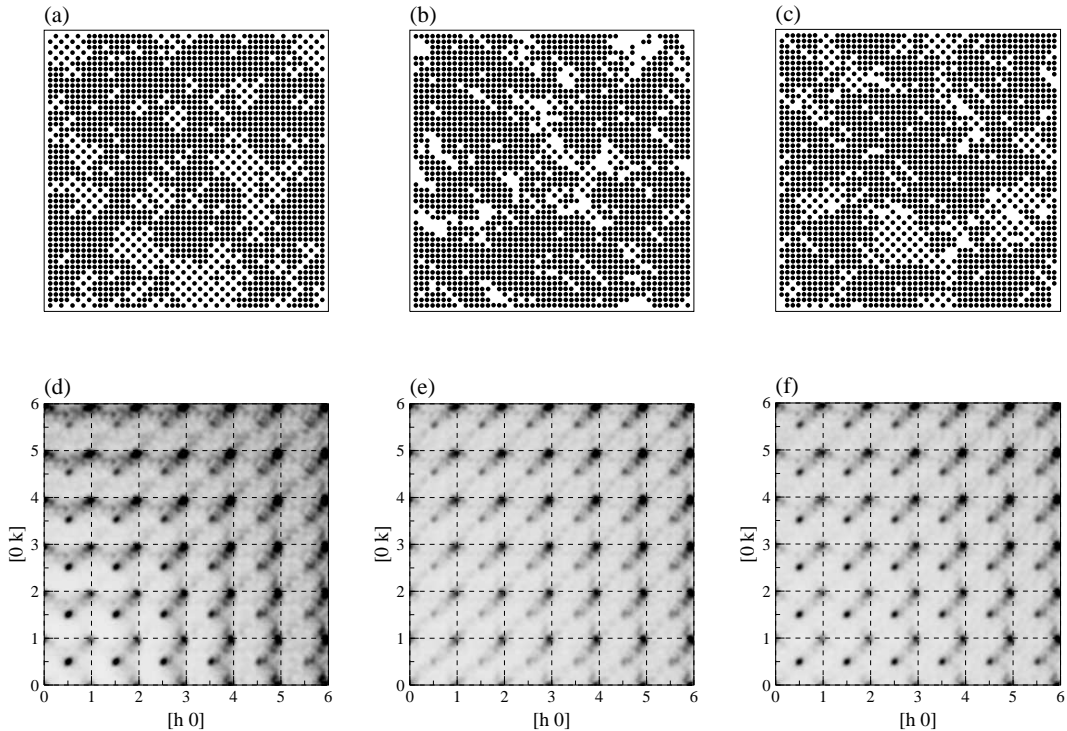}
\caption{ Input structure (a) for RMC test refinements showing preferred 
          $<$11$>$ vacancy pairs and corresponding diffraction pattern (d). 
          Resulting defect structure run A (b) and run B (c) and corresponding 
          diffraction patterns (e) and (f).}
\label{fig2}
\end{figure*}

The simulated structure and the corresponding diffraction pattern are shown 
in Figure \ref{fig2}a and \ref{fig2}d, respectively.  The scattering pattern 
was computed on a grid of 301x301 points for neutron scattering at a 
wavelength of 1\AA .  \par

The RMC simulations started from a structure showing a random vacancy 
distribution and displacements corresponding to an isotropic temperature 
factor of B=0.5 \AA$^{2}$.  Simulation A was carried out by alternately 
executing one cycle in 'swap atoms' mode followed by one cycle in 'swap 
displacements' mode.  A total of 15 cycles of each RMC mode was computed.  
In this paper a cycle will be the number of RMC moves necessary to visit 
each crystal site once on average.  The resulting correlations, 
displacements and R-values are listed in Table \ref{tab1}.  The resulting 
final distortions are consistent with the expected {\it size-effect} like 
displacements, i.e.  the atoms surrounding a vacancy are shifted towards 
the vacancy.  The calculated diffraction pattern (Fig. \ref{fig2}e) shows a 
satisfactory agreement with the experimental data at high ${\bf Q}$ values, 
where as diffuse features at positions $(\frac{h}{2}, \frac{k}{2})$, with 
$h$ and $k$ integer, at low ${\bf Q}$ are barely reproduced.  These diffuse 
intensities are mainly caused by the vacancy ordering.  The correlation 
values achieved (Tab.  \ref{tab1}) show positive values in the $<$10$>$ and 
$<$11$>$ directions rather than the expected negative correlation in 
$<$10$>$ direction.  Consequently the resulting structure (Fig. \ref{fig2}b)
shows a clustering of the vacancies rather than the expected vacancy ordering 
present in the model structure.  It appears that the dominant part of the 
diffuse scattering caused by the displacements has a too strong an 
influence on the correlations that are achieved.  \par

In order to model both parts of the defect structure more simultaneously, 
during run B the 'swap atoms' and 'swap displacements' modes were 
alternated every 0.1 cycles, i.e.  the number of moves necessary to visit 
10\% of all atom sites on average.  Additionally, the data set used for the 
'swap atoms' moves was limited to a range of $|{\bf Q}| < 4.0$ r.l.u., i.e.  
only the low angle part, less affected by diffuse scattering due to 
distortions, was used to refine the vacancy ordering.  The final values for 
the distortions (Tab.  \ref{tab1}) are similar to those of the previous 
run.  However, visual inspection of the calculated diffraction patterns 
(Fig.  \ref{fig2}f) shows a much better agreement, especially at low ${\bf Q}$ 
values, with the input data set compared to run A, although the R-value has 
only slightly improved.  The resulting structure (Fig. \ref{fig2}c) and 
correlation values achieved (Tab.  \ref{tab1}) reflect the significantly 
improved description of the simulated disordered structure.  \par

The simulation results presented here indicate that the RMC simulation 
technique is a powerful tool to analyse diffuse scattering and to obtain 
information about the defect structure even of quite complex systems 
showing both occupational and displacement disorder. However, the 
resulting correlation and displacement values are still significantly 
smaller than the expected values present in the input structure. 
Current efforts are to improve the results of RMC refinements by using 
a different way to calculate the diffuse diffraction pattern. Run C in 
Table \ref{tab1} shows the results achieved using this modified 
technique. More details can be found in \cite{prwe97d} and in
the discussion of the RMC method in section \ref{lots}.

\subsection{Stabilised zirconia}

The cubic phase of pure $ZrO_{2}$ is thermodynamically stable only at 
temperatures above 2370C.  This phase, however, can be stabilised at room 
temperature by doping with oxides of a variety of lower valent metals, e.g.  
$CaO$, $MgO$, $Y_{2}O_{3}$.  The average structure is of fluorite type, 
space group {\it Fm3m}, with zirconium on $(0,0,0)$ and oxygen on 
$(\frac{1}{4},\frac{1}{4},\frac{1}{4})$.  The dopant cation occupies a 
zirconium site and in order to maintain charge neutrality, a corresponding 
number of oxygen vacancies is introduced resulting in important ceramic and 
ionic conduction properties of these materials.  The diffuse scattering of 
these materials has been investigated by numerous workers (see references 
in \cite{prnefras93,webuthwi93}.  More recently, the authors of 
\cite{prnefr96} described the diffuse X-ray and neutron scattering of 
Ca-CSZ ($Zr_{0.85}Ca_{0.15}O_{1.85}$) by a model of correlated microdomains 
using a formula for the diffuse intensity given by \cite{nefrsc90,nefrsc90b}. 
The model consists of two types of microdomains, one based on 
a single vacancy with relaxed neighbouring atoms, the other based on a pair 
of vacancies separated by $\frac{1}{2}<$111$>$ over those oxygen cubes 
containing a metal atom.  A quite different 'modulation wave' approach to 
model the vacancy distribution by \cite{wewima95} provided further evidence 
for the existence of $\frac{1}{2}<$111$>$ vacancy pairs.  A model for the 
diffuse scattering of yttrium stabilised zirconia (Y-CSZ) with a 
composition of $Zr_{0.61}Y_{0.39}O_{1.805}$ based on Monte Carlo (MC) 
simulations was given by \cite{webuthwi93,wewithbu92} Our most recent work 
on the diffuse scattering of CSZs \cite{prwe97b} uses the RMC simulation 
technique to analyse the complex defect structure of these 
materials. \par

\begin{figure*}[!hbt]
\centering
\includegraphics[angle=270,width=5.6in]{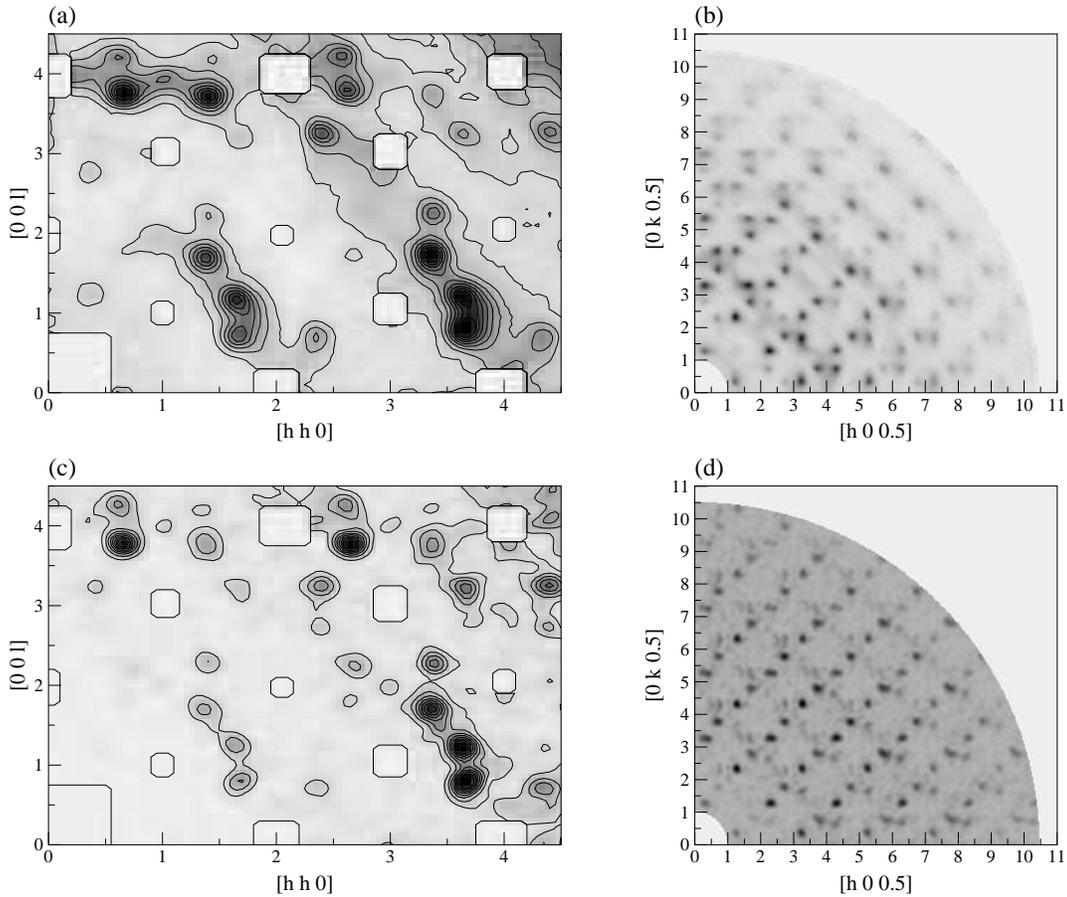}
\caption{ Diffuse scattering of Ca-CSZ: (a) neutron scattering of 0th 
          layer of the [$1\overline{1}0$]-zone and (b) X-ray scattering of 
          layer $0.5c^{*}$.  Corresponding resulting diffraction patterns 
          from RMC refinement: (c) neutron and (d) X-ray. }
\label{fig3}
\end{figure*}

These RMC simulations of Ca-CSZ used neutron as well as X-ray diffuse 
scattering data as input.  The X-ray data were collected on our PSD 
diffractometer system \cite{oswe90}.  The neutron diffraction data used in 
this study were collected by \cite{nefrsc90}.  A total of three layers of 
X-ray diffuse scattering, i.e.  $0.3c^{*}$, $0.5c^{*}$ and $0.7c^{*}$, and 
two layers of neutron diffuse scattering, i.e.  0th and 2nd layer of 
[$1\overline{1}0$]-zone, were used as input for the RMC simulations (two 
layers are shown in Fig.  \ref{fig3}a,b).  Furthermore, the cubic symmetry of
the crystal was taken into account assuming, that all symmetrically equivalent 
planes contain the same data as the experimental planes.  This resulted in 
an effective number of data points in excess of 900000.  The first RMC 
simulations were carried out using the largest computationally feasible 
crystal size of 20x20x20 unit cells containing a total of 96000 atoms 
(including vacancies).  A total of 5 RMC cycles was carried out, giving a 
R-value of 29.6\% (Run A).  However, the resulting correlations (Tab.  
\ref{tab2}) suggest that the good fit was mainly obtained by longer ranging 
correlations rather than the local disorder we are interested in.  \par

\begin{table}
\begin{tabular}{l r@{.}l r@{.}l}
  {\it Neighbour} & 
  \multicolumn{2}{c}{{\it Run A}} & 
  \multicolumn{2}{c}{{\it Run B}} \\
  \hline
  \multicolumn{5}{c}{{\bf Correlations}} \\ 
  \hline
  $VAC-VAC: \frac{1}{2} \langle 100 \rangle$     & -0&005 & -0&008(11) \\
  $VAC-VAC: \frac{1}{2} \langle 110 \rangle$     & -0&014 & -0&011(10) \\
  $VAC-VAC: \frac{1}{2} \langle 111 \rangle^{*}$ &  0&006 & -0&009(17) \\
  $VAC-VAC: \frac{1}{2} \langle 111 \rangle$     &  0&008 &  0&015(7)  \\
  $Ca-Ca:   \frac{1}{2} \langle 110 \rangle$     & -0&037 & -0&052(33) \\
  $Ca-Ca:               \langle 100 \rangle$     & -0&009 & -0&008(17) \\
  \hline
  \multicolumn{5}{c}{{\bf Displacements [\AA]}} \\
  \hline
  $VAC-O:     \frac{1}{2} \langle 100 \rangle$    & -0&011 & -0&031(13) \\
  $VAC-O:     \frac{1}{2} \langle 110 \rangle$    &  0&003 &  0&009(5)  \\
  $VAC-O:     \frac{1}{2} \langle 111 \rangle$    &  0&005 &  0&014(7)  \\
  $Zr-Zr: \frac{1}{2} \langle 110 \rangle ^{VAC}$ &  0&024 &  0&064(39) \\
  $Zr-Zr: \frac{1}{2} \langle 110 \rangle ^{O}$   & -0&002 & -0&012(6)  \\
  $Zr-Ca: \frac{1}{2} \langle 110 \rangle ^{VAC}$ &  0&020 &  0&029(35) \\
  $Zr-Ca: \frac{1}{2} \langle 110 \rangle ^{O}$   & -0&007 & -0&019(10) \\
\end{tabular}
\caption{\label{tab2}Results of the RMC refinements of the diffuse 
          scattering of Ca-CSZ}
\protect
\end{table}

In order to force the RMC process to model the diffuse scattering using 
correlations on a local scale, a series of RMC refinements, presented here, 
was carried out based on a crystal of only 5x5x5 unit cells in size.  A 
total of 6 RMC refinements were computed (Run B), each run iterated for 25 
cycles.  The oxygen-oxygen vacancy ordering and the Zr-Ca ordering as well 
as displacements for all atoms were modeled.  The simulation mode was 
similar to the one used for the test simulations described in the last 
section.  The resulting average R-value for all refinements is 33.1\%, the 
average of the resulting diffuse neutron and X-ray patterns of all RMC 
refinements are shown in Figure \ref{fig3}c and d, respectively.  \par

Two different types of correlations are listed in Table \ref{tab2}: The 
oxygen vacancy-vacancy correlations are represented by 'VAC-VAC' and 
given for nearest neighbours ($\frac{1}{2}<$100$>$), next-nearest 
neighbours ($\frac{1}{2}<$110$>$) and second-nearest neighbours over those 
oxygen cubes containing no cation ($\frac{1}{2}<$111$>^{*}$) and those 
cubes filled with a cation ($\frac{1}{2}<$111$>$).  Additionally the Ca-Ca 
occupancy correlations for nearest neighbours ($\frac{1}{2}<$110$>$) and 
next-nearest neighbours ($<$100$>$) are given.  Inspection of Table 
\ref{tab2} shows that the average of all RMC refinements using the 5x5x5 
unit cell crystal (Run B) results in an oxygen-vacancy ordering scheme 
similar other models proposed in the literature.  However, the given 
standard deviations indicate there is a large variation in the values 
obtained from the different RMC refinements, so that the actual values are 
barely significantly different from zero.  The RMC run A using the 20x20x20 
unit cell crystal shows the same general trends in the correlation values, 
although here magnitudes are generally even lower and in one case of 
opposite sign (VAC-VAC $\frac{1}{2}<$111$>^{*}$).  Despite the actual 
values being small, the trends are quite definite and reproducible.  It 
should be noted, that the larger values are obtained using a smaller 
crystal size, which is consistent with the view that the smaller system is 
less able to use long-range correlations to obtain the fit.  The Ca-Ca 
correlations show negative values for nearest and next-nearest neighbours 
for both refinements, although again these are very low in magnitude.  Such 
negative correlations indicate Ca-Ca nearest and next-nearest neighbours 
tend to be avoided, i.e.  less probable than in a random cation 
distribution.  Again the smaller model crystal size leads to slightly 
larger negative values.  \par

The displacements listed in Table \ref{tab2} are the distances from the 
average fluorite position.  The results show that the nearest neighbour 
oxygens are shifted towards the vacancy along the $<$100$>$-direction 
whereas next-nearest and second-nearest neighbouring oxygen atoms are 
moved away from the vacancy.  The metal-metal distance along 
$\frac{1}{2}<$110$>$ is shorter than in the average structure if both 
bridging oxygen sites are occupied.  If one or both of these bridging sites 
is vacant, the metals are shifted further apart.  The results listed in 
Table \ref{tab2} show no significant difference for the displacements of 
Zr-Zr pairs compared to Zr-Ca pairs.  As the correlation values, the 
refinement B shows the more significant values compared run A using the 
larger model crystal size.  However, the large standard deviation of these 
displacements indicates large statistical errors due to the small crystal 
size.  The resulting defect structure is consistent with previously 
reported models of the disorder in CSZ materials.  These calculations as 
well as as well as a study of the diffuse scattering of $TlSbOGeO_{4}$ 
\cite{wepr97} have shown, that the size of the model crystal is determined 
by two conflicting requirements: a {\it large} crystal size gives 
sufficiently smooth diffraction patterns and statistically significant 
correlation parameters, but fit is obtained by many longer-range 
correlations rather than by few short-range correlations which are the 
parameters we are interested in.  A {\it small} crystal size on the other 
hand reduces the number of variables and longer-range correlations but the 
calculated diffraction pattern is too noisy for a satisfactory fit and the 
resulting correlation values have large errors.  Further discussion of this 
problem and a way to improve the RMC results is given in section 
\ref{lots} of this paper.

\section{Discussion}

Generally the RMC simulation technique generates a disordered structure 
consistent with the observed diffuse scattering by minimising the 
difference between calculated and measured diffuse scattering patterns.  
The RMC test simulations described here have demonstrated that the RMC 
technique is a viable tool to analyse diffuse scattering and to obtain 
information about the defect structure even for quite complex systems 
showing occupational and displacement disorder.  The study of single 
crystal diffuse scattering of Ca-CSZ confirms that the RMC method is able 
to determine the characteristic features of the complex defect structure of 
a 'real' disordered system.  The combination of neutron and X-ray diffuse 
scattering data allowed the analysis to include oxygen-oxygen vacancy 
ordering as well as cation ordering and both oxygen and metal 
displacements.  The resulting defect structure shows features consistent 
with results previously reported in the literature.  \par

However, the resulting correlation parameters and displacements for
the simulations of the diffuse scattering of Ca-CSZ were
barely significantly different from zero although the trend towards the
particular defect structure was reproducible. One important problem of
the RMC simulation technique for single crystals is the size of the
model crystal, determined by two conflicting requirements. This 
'crystal size' problem will be discussed in more detail in the next 
section.

\subsection{The 'crystal size problem' \label{lots}}

RMC simulations of the diffuse scattering of CSZ presented
here as well as a study of the diffuse scattering of $TlSbOGeO_{4}$ 
\cite{wepr97} have shown, that the size of the model crystal
is determined by two conflicting requirements: a large crystal size gives 
sufficiently smooth diffraction patterns and statistically significant 
correlation parameters, but the fit is obtained by many longer-range 
correlations rather than by few short-range correlations which are the 
parameters we are interested in. A small crystal size on the other hand 
reduces the number of variables and longer-range correlations but the 
calculated diffraction pattern is too noisy for a satisfactory fit and 
the resulting correlation values have large errors. \par

\begin{figure}[!htbp]
\centering
\includegraphics[angle=270,width=2.8in]{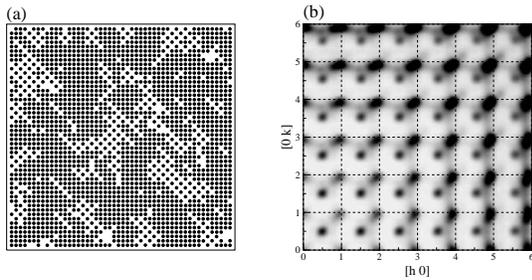}
\caption{ RMC refinement results from run C using 'lots': (a) resulting 
	  structure and (b) corresponding diffraction pattern.}
\label{fig4}
\end{figure}

One way to get around the described 'crystal size problem' is the 
calculation of the diffuse scattering pattern as the average of intensities 
calculated from small regions ('lots') within the model crystal chosen at 
random.  This procedure results in a smooth diffraction pattern but, on the 
other hand, structural changes during the RMC refinement are done on a 
local scale within a single lot.  Our current efforts are to incorporate 
this method of calculating smooth diffraction patterns by averaging the 
diffuse intensity of small areas within the crystal in the RMC refinement 
process.  Additional to the high quality diffraction patterns obtained, 
this method results in a RMC refinement using only {\it local} correlations 
within one lot.  First refinements of the diffuse scattering of the test 
structures \cite{prwe97d} described in section \ref{test} of this paper 
using 'lots' were carried out (Run C).  Although the same disordered 
structure was used for these tests, the diffuse scattering used as 
'experimental data' for the RMC refinement was calculated using 'lots' as 
well.  Obviously the size of these 'lots' must be large enough to contain 
all significant interactions.  The resulting structure and diffraction 
pattern are shown in Figure \ref{fig4}a and 4b.  Inspection of Table \ref{tab1} 
shows for Run C correlations values and displacements significantly closer to 
the expected values present in the input structure.  It should be noted, that 
the refinement using 'lots' requires substantially more computer resources 
compared to 'normal' RMC refinements.  These test simulations demonstrate 
that the new RMC simulation technique using 'lots' improves the resulting 
local disorder for the type of systems discussed in this paper. \par

One might argue, that the desired restriction to a 'local' scale in fact 
limits the potential of the RMC method.  However, for many systems a 
'local disorder' model gives usually not only the simplest description of 
the particular disorder, but the local chemistry and near-neighbour 
interactions are frequently the interesting properties of the found defect 
structure.  The test examples (see section \ref{test}) showing mainly short 
range order, were significantly better refined using the modified RMC 
simulation method using 'lots' compared to the 'normal' RMC refinement.  
One should bear in mind, that the RMC method produces the most disordered 
structure consistent with the data \cite{nikemc95}.

\subsection{What next ?}

The RMC simulation technique has a large potential for further
improvements and developments and, in the authors opinion, will continue 
to play its important role as method to analyse single crystal 
diffuse scattering. \par

Further developments of the RMC simulation technique to analyse diffuse 
scattering of single crystals are twofold: First constrains like minimal 
allowed distances between atom types can be used to avoid resulting 
structures which are unlikely or even impossible from a chemical of 
physical point of view.  Those constrains can be highly specialised for a 
particular problem.  A second major improvement can be expected from RMC 
simulations combining single crystal diffuse scattering data with other 
experimental data, e.g.  powder diffraction data or EXAFS data.  Thus a 
large RMC model could be constrained using experimental data more dependent 
on short range order.  An the other hand unwanted long range fluctuations 
of might be suppressed by including Bragg data in the RMC refinement 
process.  Besides the options already mentioned, one significant 
improvements would certainly be the possibility to used fully 3-dimensional 
data sets of diffuse scattering for the RMC refinement which is currently 
beyond available computer resources.  With the continuing increase of 
available computing power, we will certainly see further developments in 
the area of computer simulations to aid the analysis of diffuse scattering 
in general and the RMC method in particular.

\section{Acknowledgments: }

The RMC refinements of the diffuse scattering of CSZ were carried out 
on a Fujitsu VPP-300 supercomputer using a grant from the Australian National 
Supercomputer Facility. The  work was supported in part by funds of the 
DFG (grant no. Pr 527/1-1).






\end{document}